\documentclass{emulateapj}
\usepackage{apjfonts,amsmath}
\usepackage{color}
\newcommand{\minusone}{$^{-1}$}

\newcommand{\zdot}{$\dot{z}$}

\shorttitle{Toward a Direct Measurement of the Cosmic Acceleration}
\shortauthors{Darling}

\begin{document}
\title{Toward a Direct Measurement of the Cosmic Acceleration}

\author{ Jeremy Darling\altaffilmark{1}}
\altaffiltext{1}{Center for Astrophysics and Space Astronomy,
Department of Astrophysical and Planetary Sciences,
University of Colorado, 389 UCB, Boulder, CO 80309-0389; jdarling@colorado.edu}

\begin{abstract}
We present precise \ion{H}{1} 21 cm absorption line redshifts observed in multiple 
epochs to directly constrain the secular redshift drift 
$\dot{z}$ or the cosmic acceleration, $\Delta v/\Delta t_\circ$.  
A comparison of literature analog spectra to contemporary digital spectra shows significant
acceleration likely attributable to systematic instrumental errors.  
However, we obtain robust constraints using primarily Green Bank Telescope digital data.
Ten objects spanning $z=0.09$--0.69 observed over 13.5 years show
$\dot{z} = (-2.3\,\pm\,0.8)\times10^{-8}$~yr$^{-1}$ or $\Delta v/\Delta t_\circ = -5.5\,\pm\,2.2$~m~s$^{-1}$~yr$^{-1}$.
The best constraint from a single object, 3C286 at $\langle z\rangle = 0.692153275(85)$, is
$\dot{z} = (1.6\,\pm\,4.7)\times10^{-8}$~yr$^{-1}$ or $\Delta v/\Delta t_\circ =2.8\,\pm\,8.4$~m~s$^{-1}$~yr$^{-1}$.
These measurements are three orders of magnitude larger than the theoretically expected acceleration 
at $z=0.5$, $\dot{z} = 2\times10^{-11}$~yr$^{-1}$ or $\Delta v/\Delta t_\circ = 0.3$~cm~s$^{-1}$~yr$^{-1}$, 
but they demonstrate the lack of peculiar acceleration in absorption line systems and 
the long-term frequency stability of modern radio telescopes.  
A comparison of UV metal absorption lines to the 21 cm line 
improves constraints on the cosmic variation of physical constants:
$\Delta(\alpha^2g_p\,\mu)/\alpha^2g_p\,\mu = (-1.2\,\pm\,1.4)\times10^{-6}$ in the redshift
range $z=0.24$--2.04.
The linear evolution over the last 10.4 Gyr is
$(-0.2\, \pm \, 2.7) \times 10^{-16}$~yr$^{-1}$, consistent with no variation.  
The cosmic acceleration could be directly measured in $\sim$125 years 
using current telescopes or in $\sim$5 years using a Square Kilometer Array, 
but systematic effects will arise at the 1~cm~s$^{-1}$~yr$^{-1}$ level.
\end{abstract}

\section{Introduction}

The measurement of secular redshift drift can directly determine the cosmic acceleration\footnote{We use ``acceleration'' in 
the physical sense, which includes deceleration.} 
and the history of expansion in a model-independent manner \citep{sandage62,loeb98} and
has recently been explored in the context of dark energy and large aperture optical telescopes 
\citep[e.g.,][]{corasaniti07,liske08}.
The magnitude of the redshift drift is minuscule, of
order 5~cm~s$^{-1}$~decade$^{-1}$ for a galaxy at redshift $z=1$, 
similar to gravitational accelerations in galaxies and clusters \citep[e.g.,][]{amendola08}.
Measuring this effect requires extreme redshift precision 
over time baselines of decades,  and the signal may thus be overwhelmed by
other systematic effects, both physical and observational.
For example, secular redshift drift may be induced by proper acceleration of the observer.  Such 
proper acceleration (but not the associated redshift drift) has been 
very precisely measured using pulsar timing in a Galactic reference frame \citep{zakamska05} and 
in a cosmological context using the proper motion of extragalactic radio sources \citep{titov11,xu12}.
Proper acceleration of the solar system barycenter about the Galactic center creates an extragalactic dipole
proper motion secular aberration drift signature on the sky that distinguishes it from a cosmological signal:
objects appear to be moving toward the Galactic center.  
Astrometry and pulsar timing
cannot, however, detect a change in the expansion rate of the universe, which is purely radial.

\begin{figure*}
%\begin{figure}
%\epsscale{1.3}%{0.8}
\epsscale{1.0}
\plotone{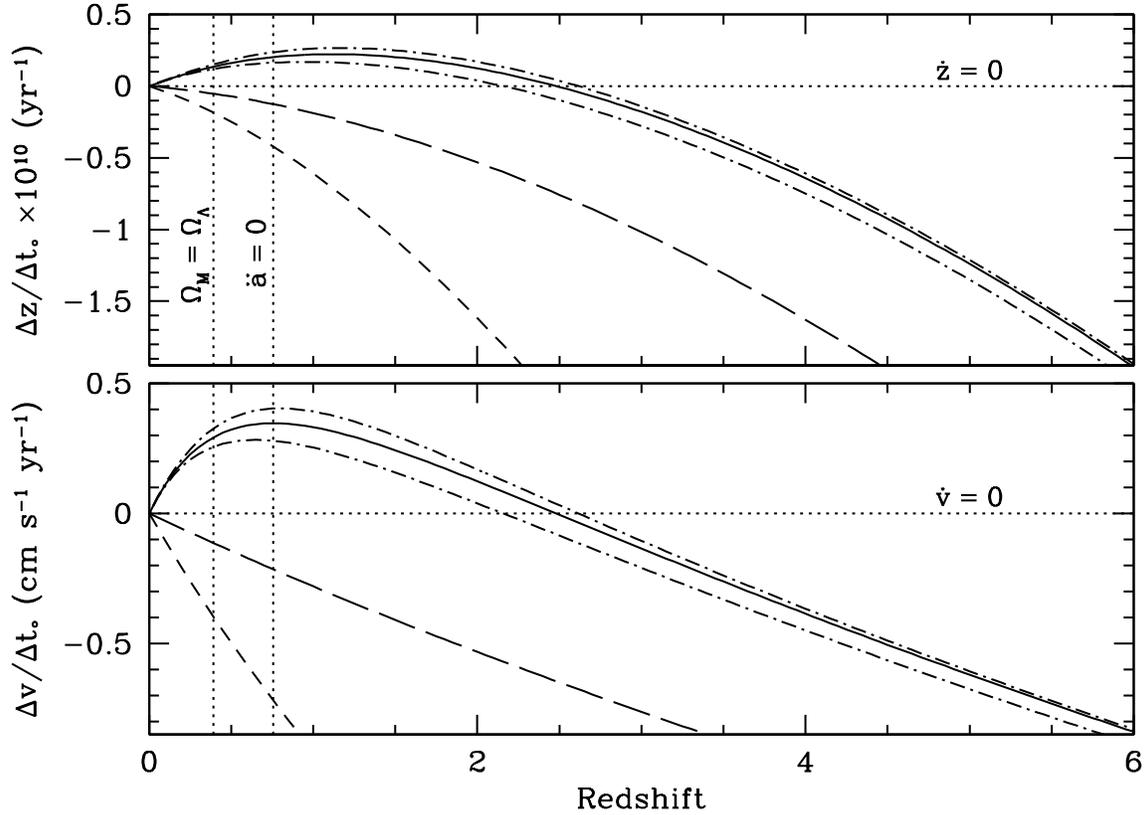}
\caption{\footnotesize
Theoretical acceleration vs.\ redshift.  
Top:  Secular redshift drift (Eqn.\ \ref{eqn:zdot}). 
Bottom:  Apparent acceleration (Eqn.\ \ref{eqn:dv}).
The solid lines show the cosmology $H_\circ = 72$~km~s$^{-1}$~Mpc$^{-1}$, 
$\Omega_\Lambda = 0.73$, and $\Omega_M = 0.27$, 
the dash-dotted lines show this cosmology with a varying dark energy equation of
state, $w(a) = w_\circ +w_a (1-a)$ \citep{linder03}, with $w_\circ=-1$ and $w_a=\pm\,0.5$ (lower and upper
curves, respectively), 
the long-dashed lines show a matter-only universe with $\Omega_M = 0.27$ (open), 
and the short-dashed lines show a closed matter-only universe ($\Omega_M = 1.0$).
\citet{balbi07} examine \zdot\ for a panoply of less conventional models.  
The horizontal dotted lines indicate stationary redshift and velocity
(crossing the concordance cosmology line at $z_{\,\dot{z}=0}=2.48$),
and the vertical dotted lines show the matter-cosmological constant
equivalence redshift ($z_{M\Lambda} = 0.39$) and the transition from a
decelerating to an accelerating universe ($z_{\,\ddot{a}=0} = 0.76$).
\label{fig:zdot_theory}}
%\end{figure}
\end{figure*}

The observed rate of change of redshift, $\Delta z/\Delta t_\circ$ (hereafter $\dot{z}\,$), is the difference
between a constant expansion rate at redshift $z$, $H_\circ (1+z)$, and its actual value $H(z)$:
\begin{equation}
  {\Delta z\over\Delta t_\circ} = H_\circ (1+z) - H(z),  \label{eqn:zdot}
\end{equation}
where 
\begin{equation}
  H(z) = H_\circ \sqrt{\Omega_{M,\circ}(1+z)^3+\Omega_\Lambda+(1-\Omega_{M,\circ}-\Omega_\Lambda)(1+z)^2}
\label{eqn:Hofz}
\end{equation}
for a matter- and cosmological constant-dominated universe, and $\Delta t_\circ$ is the 
observer's time increment \citep{loeb98}.
For a time-varying dark energy parameterized with scale factor-dependent 
equation of state $w(a) = w_\circ +w_a (1-a)$ \citep{linder03} where $a = 1/(1+z)$ is the scale
factor, the Hubble expansion becomes
\begin{multline}
 H(z) = H_\circ \sqrt{\Omega_{M,\circ}(1+z)^3+\Omega_{w,\circ}(1+z)^{3(1+w_\circ+w_a)}e^{-3 w_a z/(1+z)}}\\
 \overline{\ +\, (1-\Omega_{M,\circ}-\Omega_{w,\circ})(1+z)^2}.
\label{eqn:Hofz2}
\end{multline}
The observed acceleration is
\begin{equation}
  {\Delta v\over\Delta t_\circ} =   c\ {\Delta z/\Delta t_\circ\over(1+z)}.
\label{eqn:dv}
\end{equation}
The secular redshift drift and corresponding acceleration are plotted for several cosmologies in 
Fig.\ \ref{fig:zdot_theory}; \citet{balbi07} explore less conventional 
cosmological models and theories of gravity.  A direct measurement of \zdot\ is
completely model-independent and directly indicates the history of expansion, whatever
that may be.  

Here we examine the feasibility of 
measuring or constraining secular redshift drift using \ion{H}{1} 21~cm absorption line 
systems.  We demonstrate that with modern telescopes, observations spanning roughly a decade 
can measure redshifts to parts per 100 million and constrain the redshift drift to 
down a few m~s$^{-1}$~yr$^{-1}$.  Analog spectra show a systematic redshift offset from modern digital 
spectra, which limits the time baselines available for \zdot\ measurements and 
impacts previous constraints on the cosmic evolution of physical constants that 
employed analog 21 cm spectra.  
We revise the measurement of the evolution of $\alpha^2g_pm_e/m_p$
over the redshift span $z=0.24$--2.04.

In the following treatment of \zdot, we make no assumptions about cosmology; the only numbers 
needed are the speed of light, $c=299792.458$~km~s\minusone, and the rest frequency
of the \ion{H}{1} 21~cm spin-flip transition,  $\nu _o (HI) = 1420.405751768$~MHz, although
neither is truly needed to measure a redshift change since it is ultimately simply a shift in 
the observed frequency.

\section{Observations, Data, and Data Reduction}\label{sec:data}

\begin{figure}
%\begin{figure}[ht!]
\epsscale{1.2}
\plotone{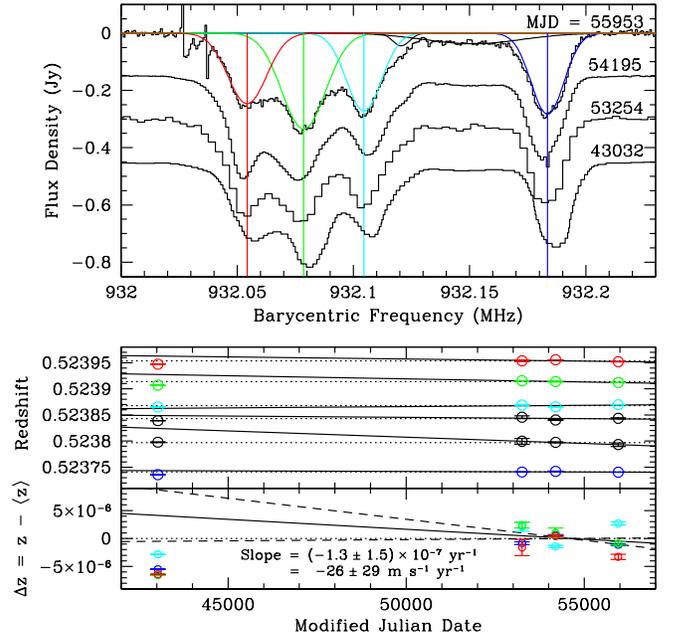}
%\begin{center}
%\includegraphics[width=1.0\textwidth]{zdot_vs_z.epsi}
%\end{center}
\caption{\footnotesize
Example of redshift drift observations of an \ion{H}{1} 21 cm absorption
line system, 0235+164.
Top:  Spectra obtained at the GBT (upper three, all digital) and Arecibo (analog, bottom; \citet{wolfe78})
labeled by MJD, scaled to match the MJD 55953 blue line, and offset for clarity.  
Gaussian fits to well-defined lines are color-coded (others
are black), matching points in the lower panel.  
As with many of the absorption line systems listed in Table \ref{tab:zdot}, the analog spectrum 
of 0235+164 is systematically offset from the digital GBT spectra and is thus omitted from the linear
fit.
Bottom:  Redshift (upper) and redshift mean offset (lower) versus MJD. 
Dotted lines indicate the no-drift (mean) locus in both panels, and solid lines show the linear fits, 
omitting the MJD 43032 analog epoch.
Dashed lines show the formal error on the slope from an error-weighted linear least-squares fit.
}\label{fig:example}
\end{figure}

Analog data provide long time baselines, in some cases more than 30 
years \citep[e.g.][]{wolfe78}, but we find a systematic offset toward lower redshifts compared to single-telescope 
digital epochs (Figure \ref{fig:example}).
These systematic offsets appear in most pre-digital spectra but are not consistent between absorption 
line systems, suggesting that they are instrumental and uncorrectable.  We thus excluded the high 
quality and potentially useful observations from analog epochs.  Moreover, most extant digital spectra lack the 
signal-to-noise needed for this study, and in order to minimize systematics we restrict all spectral epochs to those obtained
using the Green Bank Telescope\footnote{The National Radio Astronomy Observatory is a facility of the National Science Foundation operated under cooperative agreement by Associated Universities, Inc.}  (GBT), except as noted below.
 
We observed 13 \ion{H}{1} 21 cm absorption line systems in 2003--2004 
using the GBT
Spectrometer and repeated the observations in 2012 (programs GBT 03C-009 and 12A-134, respectively).
Five of the 13 sources include a third epoch obtained at the GBT (two) or elsewhere (three).
We reduced the archival GBT Spectral Processor spectra of 0235+164 and 3C286 
\citep[program GBT 07A-021;][]{wolfe11} in an identical manner to all other GBT spectra, described below.
For early third epochs (ca.\ 2000), we used 
the \citet{carilli00} European Very Long Baseline Interferometry Network spectrum of B0218+357, 
the \citet{kanekar01} Giant Metrewave Radio Telescope time-averaged spectrum of PKS~1127$-$145, and
the \citet{lane01} Westerbork Synthesis Radio Telescope spectrum of PKS~1243$-$072 (all digital observations, 
extracted by DEXTER; \citet{demleitner01}).
We observed 5 additional absorption line systems in single epochs (Table \ref{tab:zdot}).

%\begin{deluxetable}{llccccrr}
\begin{deluxetable*}{llccccrr}
%\tabletypesize{\scriptsize} 
\tablecaption{Redshift and Acceleration Measurements \label{tab:zdot}} 
\tablewidth{0pt} \tablehead{
\colhead{Absorber} & \colhead{$z_{HI}$} & \colhead{$N_{\rm Lines}$} & \colhead{$N_{\rm Obs}$} & 
\colhead{$t_\circ$} & \colhead{$\Delta t_\circ$} & \colhead{\zdot} & \colhead{$\Delta v/\Delta t_\circ$} \\
\colhead{} & \colhead{} & \colhead{} & \colhead{} & \colhead{(MJD)} & \colhead{(days)} &
\colhead{(yr$^{-1}$)} & \colhead{(m s$^{-1}$ yr$^{-1}$)}}
\startdata 
{\bf 0738+313 A}     & 0.091234959(30) & 2 & 2 & 56041 & 3053 & $-$3.0(0.9)$\times10^{-8}$ & $-$8.3(2.5) \\
{\bf 0738+313 B}     & 0.22124990(13)  & 1 & 2 & 56040 & 3050 & 2.6(3.1)$\times10^{-8}$ &  6.3(7.7) \\
PKS 0952+179 & 0.2378155(16) & 1 & 1 & 52990 &  \nodata & \nodata & \nodata \\
{\bf PKS 1413+135} & 0.24670374(30) & 2 & 2 & 55960 & 2969 &  $-$2.3(6.9)$\times10^{-8}$ & $-$5.5(16.7) \\
{\bf PKS 1127$-$145} &  0.312658286(20) & 4 & 3 & 55953 & 4573 & 1.4(1.0)$\times10^{-7}$ & 30.9(23.4) \\
{\bf 0248+430}      &   0.39408591(14)  & 5 & 2 & 55964 & 2710 & 5.8(10.6)$\times10^{-8}$ & 12.6(22.8) \\
{\bf PKS 1229$-$021} & 0.39498824(59) & 2 & 2 & 56137 & 2882 & $-$8.4(20.8)$\times10^{-8}$ & $-$18.1(44.7) \\
{\it 3C196}  &  0.43667498(93) & 2 & 2 & 56136 & 2882 & $-$8.0(2.5)$\times10^{-7}$ & $-$168(52) \\
{\it PKS 1243$-$072} &  0.4367410(14) & 1 & 3 & 56136 & 4343 & 5.8(5.1)$\times10^{-7}$ & 122(107) \\
{\bf 0235+164}      &  0.523741603(72) & 4 & 3 & 55953 & 2699 & $-$1.3(1.5)$\times10^{-7}$ &  $-$25.9(28.7) \\
PKS 1629+120 & 0.5317935(11) & 2 & 1 & 56136 & \nodata & \nodata & \nodata \\
{\bf B3 1504+377} &  0.67324197(79) & 1 & 2 & 55968 & 2978 & 2.3(2.0)$\times10^{-7}$ & 40.6(35.3) \\
{\bf B0218+357} & 0.6846808(13) & 1 & 3 &56038 & 4936 & 2.1(2.6)$\times10^{-7}$ & 37.1(46.2) \\
{\bf 3C286}            & 0.692153275(85) & 1 &  3 &  55947 & 2956 & 1.6(4.7)$\times10^{-8}$ &  2.8(8.4) \\
{\it PKS 1830$-$211} & 0.8848633(41) & 5 & 2 & 56039  & 3048 & $-$1.2(1.0)$\times10^{-6}$ & $-$192(163) \\
1331+170    &   1.7763904(66)    & 1 & 1 & 53105 & \nodata & \nodata & \nodata \\
1157+014    &   1.943670(12)      & 1 & 1 & 53070 & \nodata & \nodata & \nodata \\
0458$-$020 &   2.0393767(21) & 2 & 1 & 53071 & \nodata & \nodata & \nodata
\enddata 
\tablecomments{ For each absorption line system, 
$z_{HI}$ is the error-weighted mean redshift of the strongest component
in the barycentric reference frame, 
$N_{\rm Lines}$ is the number of components used to determine \zdot, 
$N_{\rm Obs}$ is the number of observations, 
$t_\circ$ is the Modified Julian Date (MJD) of the most recent observation, 
$\Delta t_\circ$ is the full time span of the observations, 
\zdot\ is the secular redshift drift, and
$\Delta v/\Delta t_\circ$ is the acceleration (Eqn.\ \ref{eqn:dv}).
Parenthetical values are 1$\sigma$ uncertainties.
Objects in bold are those used in the \zdot\ analysis and Figure \ref{fig:zdot}; 
those in italics are poorly constrained due to broad blended lines (3C196 and PKS~1830$-$211) or
low signal-to-noise (PKS~1243$-$072).  We include objects with a single digital observation
for posterity, and some are compared to UV absorption lines in Figure \ref{fig:zHIvszUV}.
}
\end{deluxetable*}
%\end{deluxetable}

 The two main observational epochs described here span the transition from analog to digital television.  
Consequently, absorption line systems that were observable through the analog radio frequency interference (RFI)
comb in 2003, 
particularly those at $z>1$, were no longer observable amid the digital RFI in 2012, and lines that were 
lost to RFI in 2003 at the upper end of the old analog television allocation, PKS1629+120 and 3C216, 
were detectable in 2012.  These single-epoch observations are
listed in Table \ref{tab:zdot} and are considered in our treatment of fundamental physical constants
but are not used in the analysis of \zdot.

GBT observations were made in 5-minute position-switched (nodding) scans
using two linear polarizations, 9-level sampling, 2--3~s records,
a 1--1.5~s winking calibration diode injection signal,
and a 12.5 MHz bandwidth centered on the \ion{H}{1} line.  
Typical on-source integration times were 1--2 hours.  
We subtracted the corresponding off-source record from each individual 
on-source spectrum and normalized the difference spectrum by the sky spectrum.
Each spectral record was manually examined and flagged for RFI
then averaged in time and polarization.  After polynomial baseline subtraction and Hanning 
smoothing, we typically achieved 
3--24~mJy rms noise in 1.1--1.6~km~s$^{-1}$ channels at 1300--750~MHz ($z=0.09$--0.89) 
in 2003--2004, and 
6--30~mJy rms noise in 0.35--0.61~km~s$^{-1}$ channels in 2012 over the same redshift range.
We used GBTIDL\footnote{GBTIDL (\url{http://gbtidl.nrao.edu/}) is the data reduction package produced by NRAO 
and written in the IDL language for the reduction of GBT data.}
for all GBT data reduction.

Absorption lines were fit using Gaussian profiles in each epoch and were not constrained
by other epochs.  Residual spectra were usually indistinguishable from line-free noise.
Typically, only high signal-to-noise narrow well-separated lines were used for the \zdot\ measurement, 
and Table \ref{tab:zdot} lists the number of components used in each case.
We measured the error-weighted mean redshift of the strongest Gaussian-fit component in each 
system to provide a reference redshift for posterity and to indicate the degree of precision 
attained (Table \ref{tab:zdot}).

%\begin{twocolumn}

%\begin{twocolumn}
\begin{figure*}
\epsscale{1.1}
\plotone{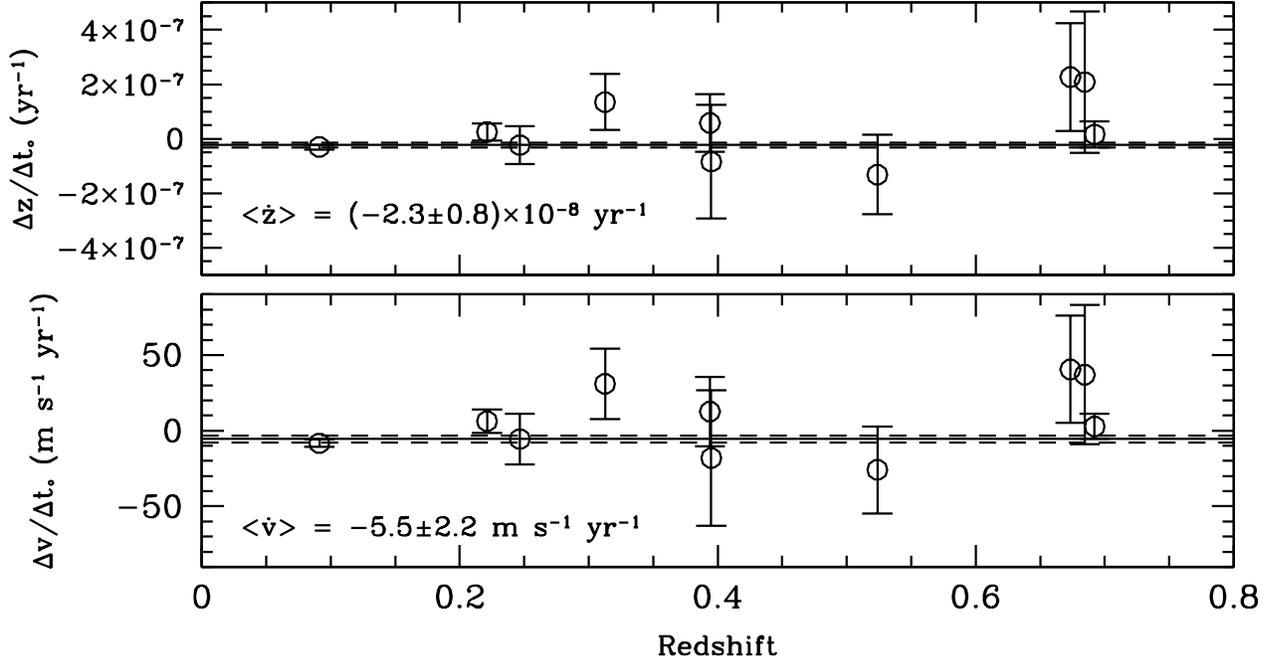}
%\begin{center}
%\includegraphics[width=1.0\textwidth]{zdot_vs_z.epsi}
%\end{center}
\caption{\footnotesize
%Example of secular redshift drifts observed in HI 21 cm absorption
%line systems.
Acceleration vs.\ redshift of \ion{H}{1} 21 cm absorption line systems.  
%[The red points include pre-digital 
%data and have long time baselines; the blue points are GBT-only observations.
Horizontal lines indicate the error-weighted mean (solid) and its 1$\sigma$ error (dashed).
The top plot shows \zdot; the bottom plot shows acceleration $\Delta v/\Delta t_\circ$.  
The error-weighted means are $\langle\dot{z}\rangle = (-2.3\,\pm\,0.8)\times10^{-8}$~yr$^{-1}$ and 
$\langle\dot{v}\rangle = -5.5\,\pm\,2.2$~m~s$^{-1}$~yr$^{-1}$, which are dominated by 
0738+313 A, 0738+313 B, PKS 1413+135, and 3C286.
}\label{fig:zdot}
\end{figure*}

Figure \ref{fig:example} shows the spectra and time series of 0235+164, 
including a pre-digital epoch, as an example of the
statistical uncertainties and systematic variation of the absorption lines.  
Weak components in the wings are only needed for good fits when the 
signal-to-noise is high, so these components are not included in the \zdot\
measurement.  We subtract the error-weighted mean redshift from each component
and then make an error-weighted linear fit to all reliable components simultaneously in order to 
average out stochastic variation among the components and obtain a bulk redshift drift of the 
absorber line ensemble.  The slope of the linear fit is \zdot.  We estimate the uncertainty in 
\zdot\ from a Monte Carlo process except for 
large $N_{\rm Lines}\times N_{\rm Obs}$ systems
0738+313~A, PKS~1127$-$145, 0248+430, and 0235+164, which use a 
formal error-weighted least-squares fitting error.
The acceleration, $\Delta v/\Delta t_\circ$, is calculated via Equation \ref{eqn:dv}
and included in Table \ref{tab:zdot}.

\section{Results and Analysis}\label{sec:results}

Figure \ref{fig:zdot} shows the secular redshift drift and acceleration of 10 absorption line
systems.  While the theoretical \zdot\ curves shown in Figure \ref{fig:zdot_theory} 
could be fit to these data and would suggest a roughly linear trend for this redshift range,
the typical measured \zdot\ value is about three orders of magnitude
larger than the theoretical expectation and is consistent with zero, so we simply calculate the mean 
\zdot\ across redshift to assess the precision of this method and the presence (or absence)
of systematic offsets from the expected value of zero.

 We measure an error-weighted mean secular redshift drift of 
$\langle\dot{z}\rangle = (-2.3\,\pm\,0.8)\times10^{-8}$~yr$^{-1}$ and an acceleration of 
$\langle\dot{v}\rangle = -5.5\,\pm\,2.2$~m~s$^{-1}$~yr$^{-1}$, both consistent with zero.  
An error-weighted linear fit fixed to $\dot{z}=0$ at $z=0$ has slope
$-3.2\, \pm\, 13.3$~m~s$^{-1}$~yr$^{-1}$ per unit redshift.
These measurements agree with expectations 
and show no evidence of secular redshift drift or other systematic effects that might 
impair efforts to obtain higher precision, either via longer time baselines, higher signal-to-noise, 
or a larger sample (Sec.\ \ref{subsec:discussion_zdot}).

Given the systematic analog-digital offsets in \ion{H}{1} 21 cm lines described in Sec.\ \ref{sec:data}, 
it stands to reason that comparisons of literature analog \ion{H}{1} lines to other absorption lines in order to 
measure physical constants at various redshifts are systematically affected.  
The redshift difference between rest-frame UV metal and \ion{H}{1} 21 cm absorption lines measures (or constrains)
the cosmic evolution of $\alpha^2\,g_p\,m_e/m_p$, where $\alpha=e^2/\hbar\,c\simeq1/137$ is the fine structure 
constant, $g_p$ is the proton gyromagnetic factor, and $m_e$ and $m_p$ are the electron and proton mass, 
respectively \citep{wolfe76}.
Since the ratio of the \ion{H}{1} 21 cm line to the UV metal lines scales as
$\alpha^2\,g_p\,\mu$, where $\mu = m_e/m_p$, the 
redshift difference between the lines can be related to fundamental constants as
\begin{equation}
 {\Delta(\alpha^2\,g_p\,\mu)\over (\alpha^2\,g_p\,\mu)_\circ} = {z_{UV}-z_{21}\over1+z_{21}}
\end{equation}
provided the gas is velocity-coincident.

\begin{figure*}
%\begin{figure}
\vspace{5pt}
\epsscale{1.0}
\plotone{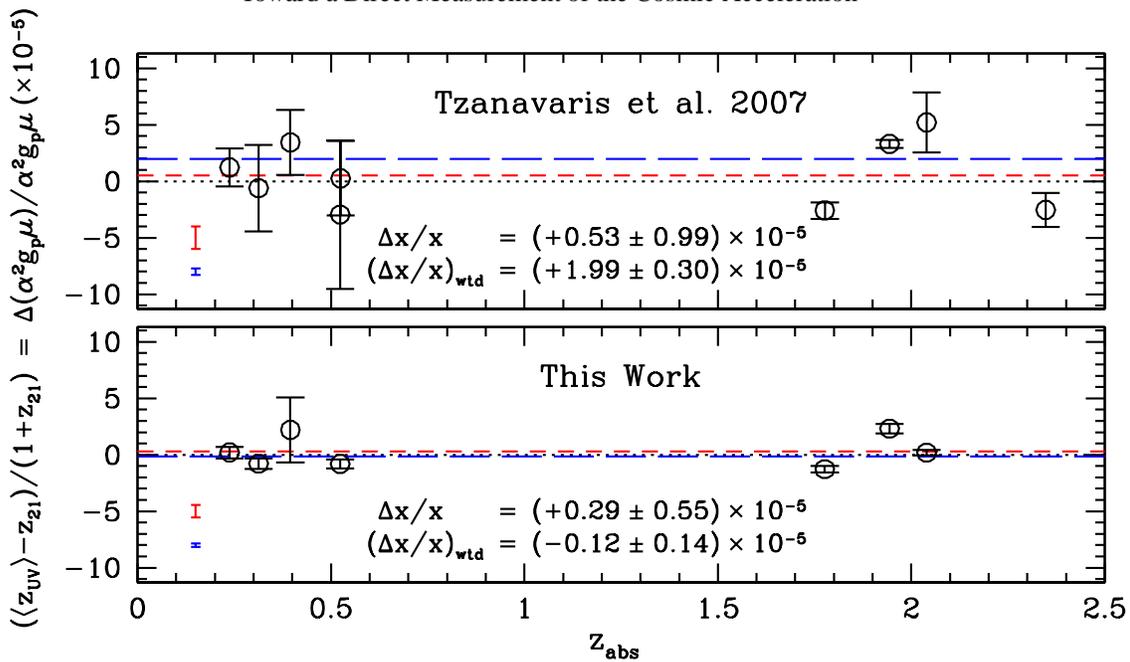}
\caption{
The fractional redshift difference between UV metal and
21 cm \ion{H}{1} absorption lines vs.\ redshift.  This is equivalent to
the fractional change in $\alpha^2\, g_p\, \mu$. %, a combination of the 
%fine structure constant $\alpha$, the proton gyromagnetic factor $g_p$, and the electron-to-proton
%mass ratio $\mu$.
Top:  Reproduction of the \citet{tzanavaris07} plot using literature 21 cm absorption lines 
and their statistical error bars.
Bottom:  the same plot using GBT-measured \ion{H}{1} 21 cm lines.
Horizontal lines indicate the null line (dotted), the mean (red, short dashed), and the error-weighted 
mean (blue, long-dashed).  Error bars in the lower left of each panel indicate the uncertainties in 
each mean value.  
Despite the smaller sample, our mean offsets are significantly smaller and their uncertainties 
are smaller by a factor of two.
Individual error bars are likewise substantially smaller due to higher signal-to-noise
in most 21 cm observations.  
}\label{fig:zHIvszUV}
\end{figure*}

\citet{tzanavaris07} employed a number of analog measurements of the 21 cm lines 
presented in Table \ref{tab:zdot} to examine the cosmic evolution of $\alpha^2\,g_p\,\mu$, so we revisit their analysis
using digital data.  
Figure \ref{fig:zHIvszUV} shows the normalized redshift difference between \ion{H}{1} 21 cm lines and UV 
metal lines for the nine absorption line systems measured by \citet{tzanavaris07} and for the seven systems
measured by us using the GBT.  The UV line ensemble redshifts are presented in \citet{tzanavaris07}.  
In two cases we use a 21 cm component that is not the dominant one for comparison
to the UV lines:  in 0235+164 we use the \ion{H}{1} line at $z=0.52384114(41)$, the strongest
of the line triplet (Fig.\ \ref{fig:example}), to match the component used by \citet{tzanavaris07}, and 
in 0458$-$020 we use the redder component at $z=2.0395474(64)$
that matches the many UV lines presented in \citet{tzanavaris07}.

%\end{figure}

 Despite the smaller sample, we obtain a substantially reduced error-weighted mean:
the fractional change in $\alpha^2\,g_p\,\mu$ is $(-1.2\pm1.4)\times10^{-6}$ 
in the redshift range $z=0.24$--2.04.
An identical treatment of the \citet{tzanavaris07} data yields a fractional change of 
$(+19.9\pm3.0)\times10^{-6}$ (using their statistical error bars for consistency).
The unweighted means are similar, 
$(+2.9\,\pm\,5.5)\times10^{-6}$ (this work) versus
$(+5.3\,\pm\,9.9)\times10^{-6}$ \citep{tzanavaris07}, and our uncertainty is roughly half 
despite the slightly smaller sample.  

The linear evolution over the last 10.2 Gyr\footnote{From present to $z=2.04$, assuming 
$H_\circ = 72$~km~s$^{-1}$~Mpc$^{-1}$, $\Omega_\Lambda = 0.73$, and $\Omega_M = 0.27$.}, 
assuming the laboratory value at $z=0$, is
$(-0.2\, \pm \, 2.7) \times 10^{-16}$~yr$^{-1}$, consistent with no variation.  A negative value indicates
a positive slope in Fig.\ \ref{fig:zHIvszUV} (the value decreases as time advances from the 
early universe to the present).    \citet{tzanavaris07} obtained a limit one order of magnitude larger, 
$(-0.6\, \pm \, 1.2) \times 10^{-15}$~yr$^{-1}$, using their favored analysis.

\section{Discussion}\label{sec:discussion}

\subsection{\boldmath{$\dot{z}$}}\label{subsec:discussion_zdot}

The intrinsic stability of the \ion{H}{1} 21 cm lines does not seem to be an issue at
the level of a few m~s$^{-1}$~yr$^{-1}$.  
Most systems show negligible epoch-to-epoch redshift variation, but a few systems such as PKS~1127$-$245
do show significant apparently stochastic variation, and the measured \zdot\ may be affected by the time of measurement
and the number of epochs.  
Stochastic line shifts of order 500~m~s$^{-1}$ over time scales of
about 1 year with no net change in the mean line velocity have been
observed (Wolfe et al.\ 1982; Briggs 1983)
as well as variation in line depth \citep[e.g.,][]{kanekar01}, and the
common interpretation is that a variable illuminating radio source structure
picks out varying portions of the foreground absorbing clouds.  This effect may become 
significant as one approaches the precision needed to detect the expected cosmological acceleration.

Gravitational accelerations of order 1~cm~s$^{-1}$~yr$^{-1}$ are likely in 
galactic and extragalactic settings, such as the acceleration of galaxies in clusters 
and the orbital accelerations within galaxies \citep{amendola08}.  For example, the Solar System 
barycenter accelerates toward the Galactic Center at $\sim$0.7 cm s$^{-1}$~yr$^{-1}$,
assuming a circular orbit.
Extragalactic gravitational accelerations are distinguishable from the cosmic \zdot\ 
because they will be randomly distributed, showing both positive and negative values, and will not be 
coherent across the sky or depend on redshift.  
The barycenter acceleration will appear in \zdot\ measurements as a redshift-independent acceleration dipole on the sky.

Our \zdot\ measurement depends primarily on the four best-measured 
absorption line systems; omitting points with error bars greater than 20~m~s$^{-1}$~yr$^{-1}$
has a negligible impact on the result, which suggests that only $\sim$10
additional systems with uncertainties of order a few m~s$^{-1}$~yr$^{-1}$ 
would improve this measurement substantially, perhaps below 1~m~s$^{-1}$~yr$^{-1}$.

To estimate the time required to measure the canonical cosmological \zdot\ at $z=1$
(\zdot~$=2\times10^{-11}$~yr$^{-1}$),
we assume a scenario whereby an ensemble of 30 lines is observed
once per year with uncertainty $\sigma_z=5\times10^{-8}$ per line per observation.  We claim a detection when
the uncertainty of a linear fit of redshift versus time recovers the slope with at least 3$\sigma$ 
significance.  For the GBT, a direct detection of \zdot\ will require about 300 years.  
For a Square Kilometer Array (SKA) with $\sim$130 times the collecting area, $\sigma_z\simeq4\times10^{-10}$ and
this measurement can be made in about 12 years, which 
is competitive with a 42~m optical telescope observing Lyman $\alpha$ forest
lines over $\sim$20 yr \citep{liske08}.  If the number of lines increases to 300, an expected dividend of upcoming 
SKA prototype surveys such as those planned for ASKAP \citep{johnston08,darling11}, 
then the GBT measurement would require $\sim$125 years, and the 
SKA measurement $\sim$5 years.

\subsection{Physical Constants}

\ion{H}{1} 21 cm lines are no longer the limiting factor in constraining the cosmic
evolution of $\alpha^2\, g_p\, \mu$:  UV metal line redshift uncertainties are 1--2 dex %orders of magnitude 
larger than the 21~cm line uncertainties at $z<1$ and are comparable at $z>1$.  Systematic effects, such as 
the coincidence of the 21~cm- and UV line-producing regions do not yet appear to be significant.
Systematic offsets between analog and digital 21 cm redshifts suggest that 
other astronomical tests of the cosmic variation of physical constants employing analog 21 cm 
data should be re-examined.  

Recent studies using modern radio line observations are consistent with no evolution in 
$\alpha^2\,g_p\,\mu$, albeit with larger uncertainties than our result:
\citet{wolfe11} obtain a 2$\sigma$ limit 
$\Delta(\alpha^2\,g_p\,\mu)/\alpha^2\,g_p\,\mu< 6\times10^{-6}$ using 3C286 alone, based on 
the identification of the observed optical absorption with a specific \ion{H}{1}-absorbing cloud, and
\citet{kanekar10} compared 21~cm to UV \ion{C}{1} absorption in two systems at $z\sim1.5$ to find
$(6.8\,\pm\,1.0_{\rm\, stat}\,\pm\,6.7_{\rm\, sys})\times10^{-6}$.  
Recent UV-\ion{H}{1} comparisons obtaining similar-uncertainty measurements to ours include
\citet{rahmani12} with $(-0.1\,\pm\,1.3)\times10^{-6}$ at $\langle z\rangle=1.36$ and 
\citet{srianand10} who find $(-1.7\,\pm\,1.7)\times10^{-6}$ in a single absorber at $z=3.17$.

\section{Conclusions}

We have obtained a precise constraint on the cosmic acceleration by directly measuring 
the real-time secular redshift drift of \ion{H}{1} 21 cm 
absorption line systems.  
We measure a redshift drift of $\langle\dot{z}\rangle = (-2.3\,\pm\,0.8)\times10^{-8}$~yr$^{-1}$ and 
an acceleration $\langle\dot{v}\rangle = -5.5\,\pm\,2.2$~m~s$^{-1}$~yr$^{-1}$, both consistent
with no acceleration.  
These measurements have not yet reached the theoretically expected acceleration, 
$\dot{z} = 2\times10^{-11}$~yr$^{-1}$ or $\Delta v/\Delta t_\circ = 0.3$~cm~s$^{-1}$~yr$^{-1}$
at $z=0.5$, but they demonstrate 
an encouraging lack of peculiar acceleration in absorption line systems and the frequency stability of 
modern radio telescopes required for repeatable measurements.  We thus find no significant systematic 
effects that would impede even more precise measurements, which may be done by expanding the sample, 
using longer time baselines, larger redshifts, and higher signal-to-noise observations, particularly 
from larger apertures.
Systematic effects will be significant below 1 cm s$^{-1}$ yr$^{-1}$, but 
direct measurements of the cosmic acceleration 
using multiple methods, wavelengths, and telescopes will likely be possible in a few decades.

\acknowledgments
We thank the GBT and NRAO staff for supporting this project and for building and operating 
an exquisitely sensitive and stable telescope.  We also thank the referee for helpful comments.
This research made use of the NASA/IPAC Extragalactic Database (NED) which is operated by the Jet Propulsion 
Laboratory, California Institute of Technology, under contract with NASA.


\begin{thebibliography}{}
\bibitem[Amendola et al.(2008)]{amendola08} Amendola, L., Balbi, A., \& Quercellini, C.\  2008, Physics Letters B, 660, 81
\bibitem[Balbi \& Quercellini(2007)]{balbi07}  Balbi, A. \& Quercellini, C.\ 2007, \mnras, 382, 1623
\bibitem[Briggs(1983)]{briggs83} Briggs, F.~H.\  1983, \apj, 274, 86
\bibitem[Carilli et al.(2000)]{carilli00}  Carilli, C.~L., Menten, K.~M., Stocke, J.~T., Perlman, E., Vermeulen, R., Briggs, F.,
    de Bruyn, A.~G., Conway, J., \&  Moore, C.~P.\  2000, \prl, 85, 5511
%\bibitem[Champion et al.(2010)]{champion10}  Champion, D. J., et al.  2010, \apj, 720, L201
\bibitem[Corasaniti et al.(2007)]{corasaniti07} Corasaniti, P.-S., Huterer, D., \& Melchiorri, A.\  2007, Phys Rev D, 75, 062001 % Sandage-Loeb test in the dark energy redshift desert, z=2--5
\bibitem[Darling et al.(2011)]{darling11}  Darling, J., Macdonald, E.~P., Haynes, M.~P., \& Giovanelli, R.  2011, \apj, 742, 60
\bibitem[Demleitner et al.(2001)]{demleitner01}  Demleitner, M., Accomazzi, A., Eichhorn, G., Grant, C.~S., Kurtz, M.~J.,
  \& Murray, S.~S.\ 2001, ASP Conf.\ Proc., Astronomical Data Analysis Software and Systems X, eds.\ F.~R. Harnden, Jr., 
  F.~A. Primini, \^ H.~E. Payne, 238, 321
\bibitem[Johnston et al.(2008)]{johnston08}  Johnston, S., Taylor, R., Bailes, M., et al.  2008, Experimental Astronomy, 22, 151
\bibitem[Kanekar \& Chengalur(2001)]{kanekar01} Kanekar, N. \& Chengalur, J.~N.  2001, \mnras, 325, 631
\bibitem[Kanekar et al.(2010)]{kanekar10}  Kanekar, N., Prochaska, J.~X., Ellison, S. L., \& Chengalur, J.~N.  
  2010, \apj, 712, L148
\bibitem[Lane \& Briggs(2001)]{lane01}  Lane, W.~M. \& Briggs, F.~H.  2001, \apj, 561, L27
\bibitem[Linder(2003)]{linder03}  Linder, E.~V.  2003, \prl, 90, 091301
\bibitem[Liske et al.(2008)]{liske08} Liske, J., Grazian, A., Vanzella, E., et al.\ 2008, MNRAS, 386, 1192 % zdot with ELTs; detailed sims of obs
\bibitem[Loeb(1998)]{loeb98} Loeb, A.  1998, ApJ, 499, L111
\bibitem[Rahmani et al.(2012)]{rahmani12}  Rahmani, H., Srianand, R., Gupta, N., Petitjean, P., Noterdaeme, P., \& V\'{a}squez, D.~A.\ 
2012, \mnras, 425, 556
\bibitem[Sandage(1962)]{sandage62} Sandage, A.  1962, ApJ, 136, 319
\bibitem[Srianand et al.(2010)]{srianand10}  Srianand, R., Gupta, N., Petitjean, P., Noterdaeme, P., \& Ledoux, C.  
  2010, \mnras, 405, 1888
\bibitem[Titov et al.(2011)]{titov11} Titov, O., Lambert, S.~B., \& Gontier, A.-M.  2011, \aap, 529, A91
\bibitem[Tzanavaris et al.(2007)]{tzanavaris07} Tzanavaris, P., Murphy, M.~T., Webb, J.~K., Flambaum, V.~V., \& Curran, S.~J.\ 2007, \mnras, 374, 634
\bibitem[Webb et al.(2011)]{webb11}  Webb, J. K., King, J. A., Murphy, M. T., Flambaum, V. V., Carswell, R. F., \& Bainbridge, M. B.  2011, \prl, 107, 191101
\bibitem[Wolfe et al.(1976)]{wolfe76} Wolfe, A.~M., Brown, R.~L., \& Roberts, M.~S.\  1976, \prl, 37, 179
\bibitem[Wolfe et al.(1978)]{wolfe78} Wolfe, A.~M., Broderick, J.~J., Johnston, K.~J., \& Condon, J.~J.\  1978, \apj, 222, 752
\bibitem[Wolfe et al.(1982)]{wolfe82} Wolfe, A.~M., Davis, M.~M., \& Briggs, F.~H.\  1982, \apj, 259, 495
\bibitem[Wolfe et al.(2011)]{wolfe11} Wolfe, A.~M., Jorgenson, R.~A., Robishaw, T., Heiles, C., \& Prochaska, J.~X.  2011, 
  \apj, 733, 24
\bibitem[Xu et al.(2012)]{xu12}  Xu, M.~H., Wang, G.~L., \& Zhao, M.  2012, \aap, 544, A135
\bibitem[Zakamska \& Tremaine(2005)]{zakamska05}  Zakamska, N. L. \& Tremaine, S.\ 2005, \aj, 130, 1939  
\end{thebibliography}
\end{document}